\documentclass[%
 reprint,
superscriptaddress,
 amsmath,amssymb,
 aps,
pra,
]{revtex4-2}

\usepackage{graphicx}
\usepackage{dcolumn}
\usepackage{bm}
\usepackage{hyperref}

\begin{document}


\title{Fully-Passive Measurement-Device-Independent Quantum Key Distribution}

\author{Jinjie Li}
\affiliation{%
 Department of Physics, University of Hong Kong, Pokfulam Road, Hong Kong.
}%

\author{Wenyuan Wang}%
 \email{wenyuanw@hku.hk}
\affiliation{%
	Department of Physics, University of Hong Kong, Pokfulam Road, Hong Kong.
}%
\affiliation{HK Institute of Quantum Science $\&$ Technology, University of Hong Kong, Pokfulam Road, Hong Kong.}
\author{Hoi-Kwong Lo}
 \email{hklo@ece.utoronto.ca}
\affiliation{
 Dept. of Electrical $\&$ Computer Engineering, University of Toronto, Toronto, Ontario, M5S 3G4, Canada.
}%
\affiliation{Centre for Quantum Information and Quantum Control (CQIQC), University of Toronto, Toronto, Ontario, M5S 1A7, Canada.
}%
\affiliation{Dept. of Physics, University of Toronto, Toronto, Ontario, M5S 1A7, Canada.
}%
\affiliation{Quantum Bridge Technologies, Inc., 100 College Street, Toronto, ON M5G 1L5, Canada
}%

\date{\today}

\begin{abstract}

A recently proposed fully passive QKD removes all source modulator side channels \cite{Wang2023a}. 
In this work, we combine the fully passive sources with MDI-QKD to remove simultaneously side channels from source modulators and detectors. 
We show a numerical simulation of the passive MDI-QKD, and we obtain an acceptable key rate while getting much better implementation security, as well as ease of implementation, compared with a recently proposed fully passive TF-QKD \cite{Wang2023}, paving the way towards more secure and practical QKD systems.
We have proved that a fully passive protocol is compatible with MDI-QKD and we also proposed a novel idea that could improve the sifting efficiency.
\end{abstract}

\maketitle


\section{INTRODUCTION}
The laws of Quantum Mechanics ensure theoretically secure communication using Quantum Key Distribution (QKD) between two parties \cite{Bennett2014,Ekert1991}. 
However, practically implementing QKD systems is still a challenge due to the lack of perfect equipment \cite{Gisin2006,Tamaki2016,Bourassa2022,Yoshino2018,Lydersen2010}. 
Hence, physicists have been working on eliminating side channels arising from these imperfections. 
The concept of Measurement-Device-Independent Quantum Key Distribution (MDI-QKD),
introduced in 2012, is automatically immune to attacks on detectors. In other words,
it completely removes the risk of side channels at detectors
 \cite{Lo2012}.

To perform MDI-QKD, the two verified users, Alice and Bob, desire to communicate and send signals to a third party, Charlie.
Charlie can even be untrusted, and he should perform a Bell State Measurement and publicly announce the successful outcomes.
Decoy state analysis technique \cite{Lo2005,Hwang2003,Wang2005} could be used when there are no available single-photon sources.
A review of MDI-QKD can be found in Appendix \ref{MDI-QKD} \cite{Lo2012}.

A fully passive QKD protocol proposed in 2022 \cite{Wang2023a,Zapatero2023} offers a way to remove side channels from the source modulator, so that, for example, Trojan-horse attack or pattern effect can be prevented \cite{Gisin2006,Yoshino2018}. 
In the recently proposed fully passive QKD, the fully passive source setup is essentially a combination of passive decoy-state and passive encoding setup \cite{Wang2023a}.
Passive decoy state \cite{Curty2009,Curty2010} uses a 50:50 beam splitter (BS) to interfere with two incoming signals, the intensity of the output signal is determined by the phase difference of the incoming signals while maintaining a randomized global phase.
In the passive encoding setup \cite{Curty2010}, a polarizing beam splitter (PBS) is used. 
The polarization of the output signal is determined by the phase difference of the two incoming signals, again with a randomized global phase. 

The fully passive QKD source uses four sources with random phases, the pair-wise phase difference determines the intensities of the two arms. 
Subsequently, the phase and intensity difference of the two arms would yield an output with random polarization.
This output can be represented by a state on a Bloch sphere \cite{Wang2023a}.
The signals coming out of the fully passive source can then be post-selected by users to perform QKD \cite{Wang2023a}. 

It is a natural progression to consider combining MDI-QKD with passive QKD, in order to ensure that both detector and source modulator sides can simultaneously avoid side channels. 
This notion inspired this work, which we have named passive MDI-QKD. 
This work mainly entails of the following key contributions: 
(1) In this work, we generalized the fully passive source proposed in our previous work \cite{Wang2023a} to MDI-QKD. Since MDI-QKD involves single-photon pairs, it is non-trivial to apply the decoy-state analysis based on passively prepared mixed states to MDI-QKD, we presented a proof to show that decoy-state analysis is applicable to fully passive MDI-QKD.
(2) We have developed a new channel model for fully passive MDI-QKD that caters for arbitrary source state polarizations on a Bloch sphere (i.e. perfect polarizations undergoing arbitrary 3D rotations). Such a model has never been explicitly studied in previous works for active MDI-QKD, which usually consider perfectly prepared signal states and only 2D rotations of polarizations on the X-Z plane caused by channel misalignment.
(3) The main challenge in simulating the channel statistics was efficiently calculating high-dimensional integrations, such as Equation \ref{eq.expectation}, we have used a high-speed numerical integration library \cite{Hahn2005} to implement efficient simulations.
Due to the double sifting nature of MDI-QKD, it is inevitable to work with very small numbers throughout the project, additionally, the higher-dimensional integrations cost significant computational power, which brings additional technical challenges for numerical simulation and key rate optimization.
(4) To address the relatively low sifting efficiency, we have proposed a novel sifting method that can further improve the key rate.

In the work that originally proposed this idea, this fully passive source was applied to a BB84 protocol \cite{Wang2023a}.
Most recently, a fully passive Twin-Field (TF) QKD protocol has also been proposed, which can also remove side channels from both modulators and detectors \cite{Wang2023}. 
However, TF-QKD is significantly more challenging experimentally compared to MDI-QKD due to the requirement of remote frequency stabilization. 
The same applies to passive TF-QKD, while the requirement on frequency stability will be much less stringent for MDI-QKD which we discuss in this work.

In this paper, the details of the passive MDI-QKD protocol will be discussed in Section 2, including the passive sources, channel model, decoy state analysis, and key rate calculation.
In Section 3, a simulation result is shown with some interpretation.
In Section 4, we provide some discussion and propose a novel idea to improve the key rate.

\section{THE PROTOCOL}

In this Section, there are four subsections that cover the discussion on the fully passive sources and detectors,  post-selection, channel model, decoy state analysis, and key rate calculation.

\begin{figure}[h]
	\centering
	\includegraphics[height=7cm]{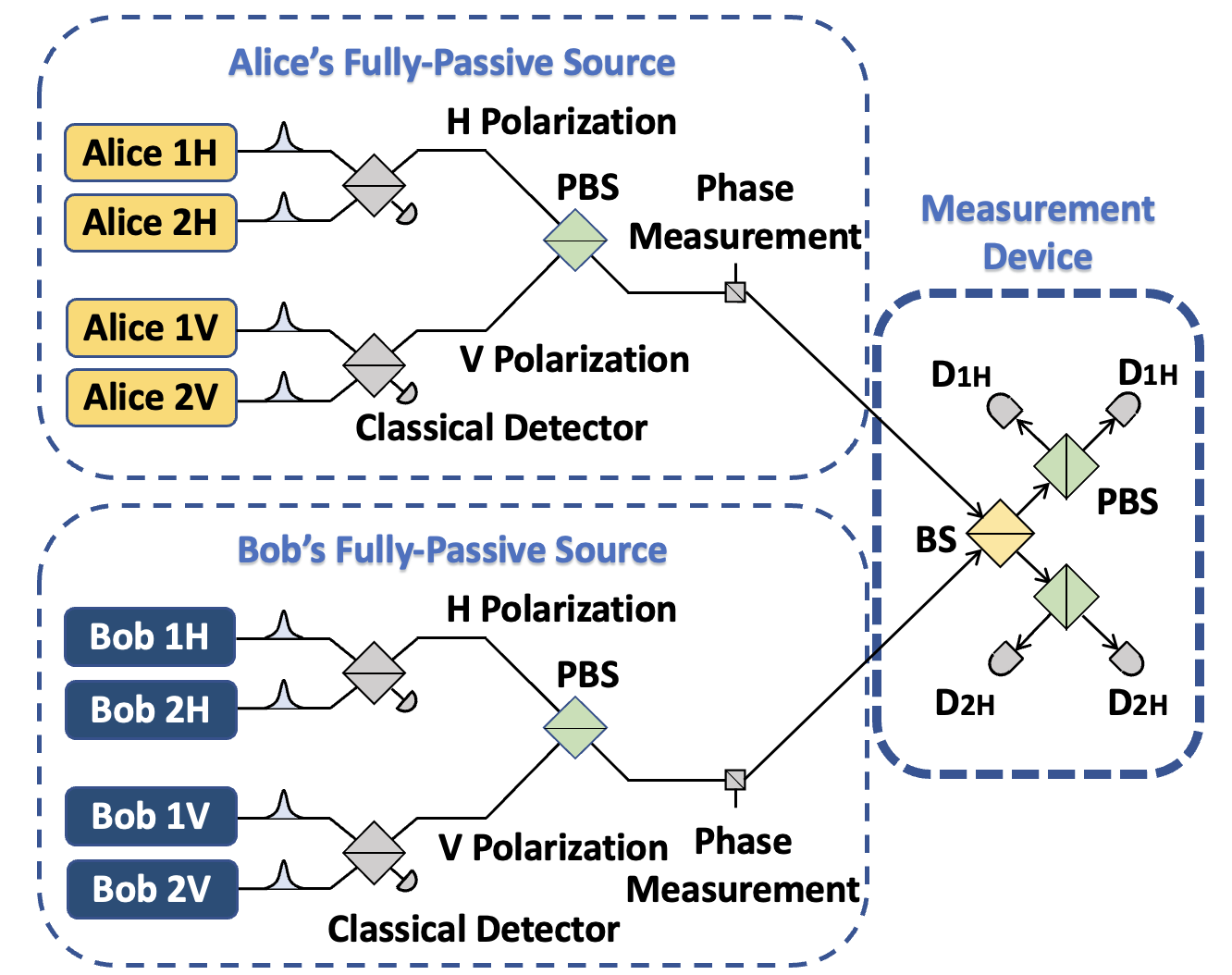}
	\caption{fully passive MDI-QKD setup. 
		Alice and Bob each possess a fully passive source, and Charlie possesses the usual MDI measurement device. 
		The fully passive sources and Charlie's setup are reproduced from \cite{Wang2023a} and \cite{Lo2012}, respectively.
	}
	\label{setup}
\end{figure}

\subsection{FULLY-PASSIVE SOURCES AND DETECTION}
A fully passive source setup \cite{Wang2023a,Lo2012} is shown in Figure \ref{setup}, Alice and Bob each hold one fully passive setup to perform passive MDI-QKD.
As shown in Figure \ref{setup}, each of Alice and Bob needs four light sources, and they join in pairs at a 50:50 BS to yield the signals at the H and V arms. 
The two arms then combine at a PBS to generate the final output signal, whose polarization is fully passively chosen \cite{Wang2023a}.
The phases of the four sources, $\phi_1,\phi_2,\phi_3,\phi_4$,  are entirely random, hence each user has four degrees of freedom (DOFs).
While Alice (or Bob) makes measurements of intensities and phases of the light signals at H and V arms, represented as $\mu_H, \mu_V, \phi_H, \phi_V$, they correspond one-to-one with a Bloch sphere coordinate $\mu, \theta_{HV}, \phi_{HV}$ and a global phase via the formula \cite{Wang2023a}

\begin{equation}
	\begin{aligned}
		\theta_{HV} & = 2 cos^{-1} (\sqrt{\frac{\mu_H}{\mu_H+\mu_V}})\\
		\phi_{HV} & = \phi_V  - \phi_H
	\end{aligned}
\end{equation}
So the 4 DOFs that Alice possesses convert from the beginning four phases to the Bloch sphere coordinates
\begin{equation}
	\phi_1,\phi_2,\phi_3,\phi_4 \rightarrow
	\mu_H, \mu_V, \phi_H, \phi_V \rightarrow
	\mu, \theta_{HV}, \phi_{HV}, \phi_{global}
\end{equation}

Since all of the four phases at the beginning are randomly chosen, the final signal can be any state on the Bloch sphere.
The output states are then ready for users, Alice and Bob, to post-select. 

Charlie, the untrusted third party, possesses the usual MDI-QKD measurement devices, see Figure \ref{setup} \cite{Lo2012}.
The four detectors Charlie uses are single photon detectors (SPDs).

\subsection{POST-SELECTION}
\begin{figure}[h]
	\centering
	\includegraphics[height=5cm]{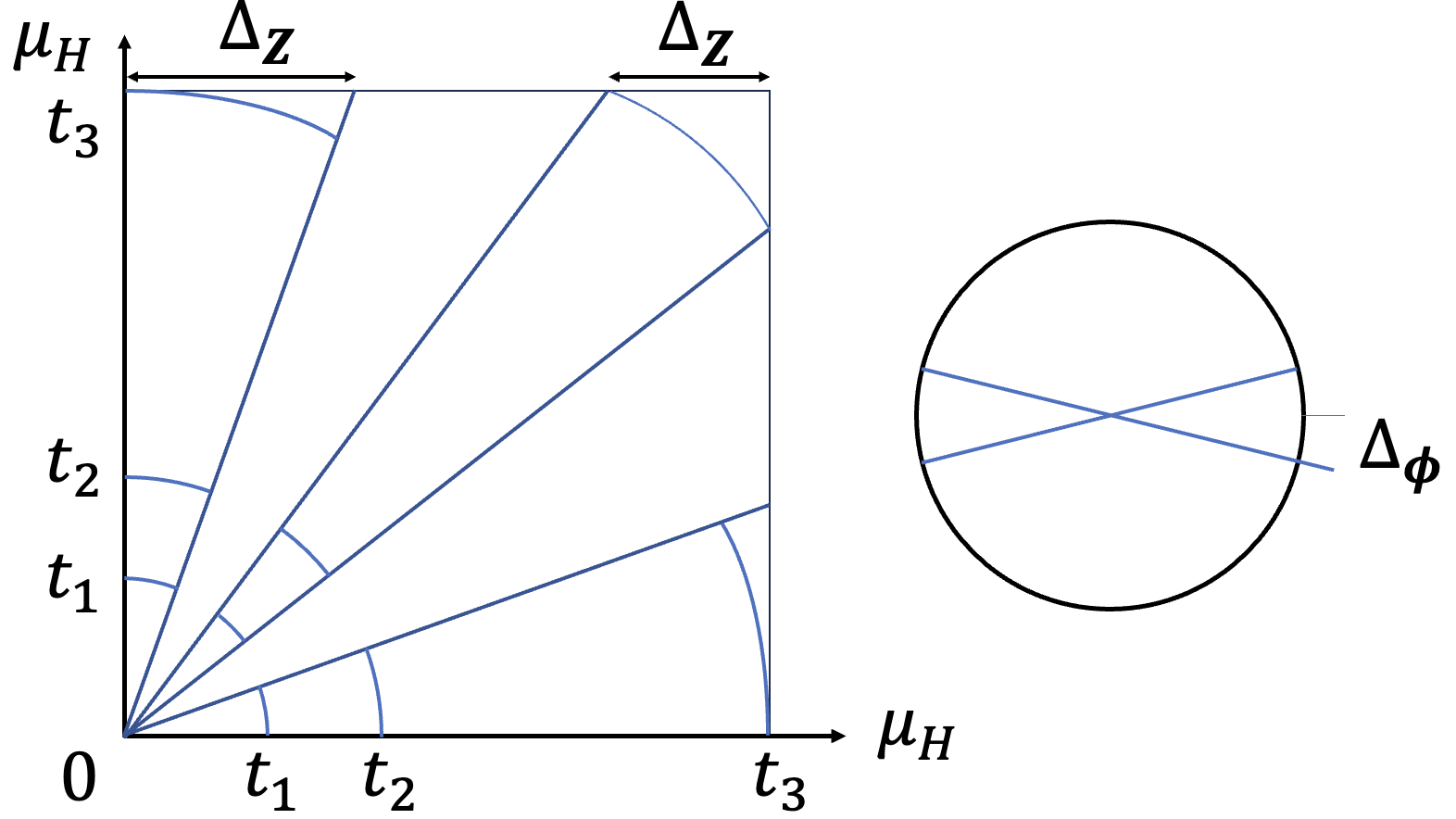}
	\caption{Alice's post-selection regions and decoy settings shown in ${\mu_H, \mu_V, \phi}$ space.
		Three decoys are used. 
		The whole setup is determined by six parameters, $\{\Delta_Z, \Delta_{XY},\Delta_{\phi}, t_1,t_2,t_3 \}$.
		Bob possesses the same decoy setting.
		This figure is reproduced from  \cite{Wang2023a}.
	}
	\label{post}
\end{figure}

The output signals from the sources can be any state on a Bloch sphere.
The users determine regions on the Bloch sphere to represent the four states $H, V, +,  -$. 
Alice and Bob post-select only those signals that lie within those regions. 
More specifically, the regions in $(\mu_H, \mu_V, \phi_{HV})$ space are shown in Figure \ref{post}.
One could subdivide the regions into smaller ones to perform decoy state analysis \cite{Wang2023a}, This point will be discussed further in Section 2.4.

Within a given region, $S_i$, there is a mixture of Bloch sphere states. 
Hence, observables are described by their expectation values, for example, for an observable $A$, its expectation value (in a fully passive MDI-QKD protocol) within given regions, $S_i$ for Alice and $S_j$ for Bob is given by

\begin{equation}
	\begin{aligned}
		\langle A\rangle_{S_{i}S_{j}} &=\frac{1}{2}\left(1 / P_{S_{i}S_{j}}\right)
		\\& \times
		 \int^{2\pi}_{0,\phi_R'} \iiint_{S_{i}} \iiint_{ S_{j}} p_A\left(\mu_{HA}, \mu_{VA}, \phi_{H VA}\right)
		 		\\& \times
		 p_B\left(\mu_{HB}, \mu_{VB}, \phi_{H VB}\right) \\
		& \times A\left(\mu_{HA}, \mu_{VA}, \phi_{H VA} ,\mu_{HB}, \mu_{VB}, \phi_{H VB} \right)
				\\& \times
		 d \mu_{HA} d \mu_{VA} d \phi_{H VA}d \mu_{HB} d \mu_{VB} d \phi_{H VB}d\phi_R', \\
		P_{S_{i}S_{j}} &=\iiint_{S_{i}} \iiint_{ S_{j}} p_A\left(\mu_{HA}, \mu_{VA}, \phi_{H VA}\right)
				\\& \times
		p_B\left(\mu_{HB}, \mu_{VB}, \phi_{H VB}\right) 
		\\
		&\times 
		d \mu_{HA} d \mu_{VA} d \phi_{H VA}d \mu_{HB} d \mu_{VB} d \phi_{H VB}.
	\end{aligned}
	\label{eq.expectation}
\end{equation}

Here, the integration of phase, $\phi_R'$ will be discussed in the next section.
The probability distribution $p_A$ for Alice is simply a classical probability, given by \cite{Wang2023a}
\begin{equation}
	p_A\left(\mu_{HA}, \mu_{VA}, \phi_{H VA}\right) = p_{\mu A}\left(\mu_{HA}, \mu_{VA}\right)p_{\phi A}\left( \phi_{HA V}\right)
\end{equation}
\begin{equation}
	\label{keydistribution}
	\begin{aligned}
		p_{\mu A}\left(\mu_{HA}, \mu_{VA}\right) &=\frac{1}{\pi^{2} \sqrt{\mu_{HA}\left(\mu_{\max }-\mu_{HA}\right)}}
				\\& \times
		\frac{1}{\sqrt{ \mu_{VA}\left(\mu_{\max }-\mu_{VA}\right)}}, \\
		p_{\phi A}\left(\phi_{H VA}\right) &=\frac{1}{2 \pi},
	\end{aligned}
\end{equation}
and Bob has the same distribution. 
Therefore,  for one observable in fully passive MDI-QKD, a 7-dimensional integration is required (three for Alice and Bob each, with an additional dimension on phase randomization), which consumes significant computational power. 
Throughout the project, all high-dimensional integrations were performed using the CUBA library \cite{Hahn2005}.

In our protocol, Alice and Bob need to perform an additional post-selection, whose probability distribution depends on the intensities for Alice or Bob

\begin{equation}
	q_{\mu}(\mu_{HA/B} , \mu_{VA/B}).
\end{equation}

This means Alice and Bob discard some of the signals according to $q_{\mu}$.
In this case, the overall probability distribution used to calculate expectation value is the product of $p$ and $q_{\mu}$.
In this work, $q_{\mu}$ is chosen to be \cite{Wang2023a}

\begin{equation}
		\begin{aligned}
			q_{\mu}&(\mu_{HA/B} , \mu_{VA/B}).
			=C \pi^{2} \sqrt{\mu_{HA/B}\left(\mu_{\max }-\mu_{HA/B}\right)}
					\\& \times
			 \sqrt{\mu_{VA/B}\left(\mu_{\max }-\mu_{VA/B}\right)} e^{\left(\mu_{HA/B}+\mu_{VA/B}\right)}
		\end{aligned}
\end{equation}

so the overall probability distribution is an exponential 

\begin{equation}
	p_{\mu} = C e^{\mu_H + \mu_V}
\end{equation}

One may find this 'modulation' of post-selection useful while doing decoy state analysis.
This will be covered in detail while discussing decoy state analysis.

\subsection{CHANNEL MODEL}

In active MDI-QKD, the two bases(X and Z basis) consists of a two-dimensional plane, while in the fully passive case, signals are represented on a three-dimensional Bloch sphere.
This poses additional difficulties when simulating its channel model.

The signals coming from Alice and Bob's fully passive setup would go through the fiber channel to reach Charlie.
Alice and Bob may prepare the states with slight misalignment.
At the same time, the signals may gain some rotation while travelling through the channels. 
A model was built to describe the misalignment and rotations based on the 3D Rodrigues' formula \cite{Friedberg2022}. 

The two signals arriving at Charlie from Alice and Bob interfere at a 50:50 BS, described by the interference formula.
Therefore, we could calculate the probability of a detection event happening at each of the four detectors.
The probabilities then can be used to calculate the Gains, $G$, and Error-gains, $QE$, given the set of input parameters, $\{\mu_{HA}, \mu_{VA}, \phi_{H VA}, \mu_{HB}, \mu_{VB}, \phi_{H VB}\}$, which describes the states coming from Alice and Bob's fully passive source.
A detailed mathematical explanation of the channel model is shown in Appendix \ref{Channel Model}.

\subsection{DECOY STATE ANALYSIS}

In QKD, the conventional decoy state analysis solves a set of linear equations to find the lower/upper bounds of single-photon yield, $Y_{11}$, and error-yield, $e_{11}Y_{11}$, for example, in MDI-QKD \cite{Xu2013}, 

\begin{equation}
	\begin{aligned}
		Q_{\mu_A  \mu_B} 
		&=\sum_{n,m=0}^{\infty}
		P_n^A P_m^B 
		Y^{nm}_{\mu_A  \mu_B}
		\\
		Q_{\mu_A  \mu_B}E_{\mu_A  \mu_B}  
		&=\sum_{n,m=0}^{\infty}
		P_n^A P_m^B 
		Y^{nm}_{\mu_A  \mu_B}e^{nm}_{\mu_A  \mu_B}
	\end{aligned}
\end{equation}
where $P$ is the Possion distribution, and $\mu_A , \mu_B$ are Alice and Bob's decoy intensity choices.

While in the passive MDI-QKD case, we have the Gains and QBERs in the form 
\begin{equation}
	\begin{aligned}
		\langle Q\rangle_{S_i^A S_j^B}
		&=\sum_{n,m=0}^{\infty}
		\left\langle P_n^A P_m^B 
		Y_{nm}\right\rangle_{S_i^A S_j^B}
		\\
		\langle QE\rangle_{S_i^A S_j^B}
		&=\sum_{n,m=0}^{\infty}
		\left\langle P_n^A P_m^B 
		e_{nm}Y_{nm}\right\rangle_{S_i^A S_j^B},
	\end{aligned}
\end{equation}
where the expectation value of any observable in passive MDI-QKD is described in Equation \ref{eq.expectation}. 

One could notice that the single-photon yield, $Y_{11}$, or the error yield, $e_{11}Y_{11}$, are within the whole seven-fold integration, hence cannot be directly 'decoupled' to a linear programming form \cite{Xu2013}.

It is important to find a way to transform this ‘coupled’ form into a ‘decoupled’ form, so that one could apply linear program (which is what is usually used in decoy state analysis to find the single-photon parameters bounds), and subsequently, it is also important to prove that the bounds derived from the ‘decoupled’ form can be used to calculate the key rate.
In the ‘decoupling’ process, we have proposed a novel post-selection method that could effectively decouple the parameters, The details of the derivation are shown in Appendix \ref{Decoy State Analysis -- Linear Programming Construction} \cite{Wang2023a}. 
The 'decoupled' forms are

\begin{equation}
	\langle Q\rangle_{S_i^A S_j^B}
	= 
	\left\langle P_n^A \right\rangle_{S_i^A}
	\left\langle P_n^B \right\rangle_{S_j^B}
	\times
	Y^{mixed}_{nm},
	\label{eq.decoy1}
\end{equation}
\begin{equation}
	\langle QE\rangle_{S_i^A S_j^B}
	= 
	\left\langle P_n^A \right\rangle_{S_i^A}
	\left\langle P_n^B \right\rangle_{S_j^B}
	\times
	e_{nm}Y^{mixed}_{nm}.
	\label{eq.decoy2}
\end{equation}

Though linear programs can be applied to Equation \ref{eq.decoy1} and \ref{eq.decoy2} to bound the single-photon yield $Y_{11}$ and error-yield $e_{11}Y_{11}$, one might notice that in Equation \ref{eq.decoy1} and \ref{eq.decoy2}, the yield is 'mixed', that is, a mixture of signals that are slightly polarized \cite{Wang2023a}. 
However, in key rate calculation, the 'perfectly prepared single photon' yield and error yield are required.
From here, we denote them $Y_{11}^{mixed}$ and $Y_{11}^{perfect}$.
We argued that the bounds for the 'mixed' quantities are actually the bounds for the 'perfectly prepared' ones by writing explicitly and making comparisons to their density matrices \cite{Wang2023a}.

For single-photon yield, $Y_{11}$, we could find that the lower bound of the mixed single-photon yield term is actually the lower bound of the 'perfect encoding state' \cite{Wang2023a}.
Mathematically, 
\begin{equation}
	Y^{mixed, Lower}_{11} \le Y^{mixed}_{11} = Y_{11}^{perfect}
\end{equation}
More specifically, we could prove that the mixed single-photon yield is actually equal to the perfectly encoded one. 
Therefore, lower bounds solved from linear programming could be used to calculate the key rate.

For single-photon error-yield, $e_{11}Y_{11}$,  the upper bound of the mixed error-yield, $e_{11}Y_{11}^{mixed}$, is actually the upper bound of the 'perfect encoding state' error yield, $e_{11}Y_{11}^{perfect}$ \cite{Wang2023a},
mathematically, 
\begin{equation}
	e_{11}Y_{11}^{mixed, Upper} \ge e_{11}Y_{11}^{ mixed}  \ge   e_{11}Y_{11}^{perfect}.
\end{equation}

In this way, we could apply the linear programming technique to find the lower bound of single-photon yield and upper bound of single-photon error rate and, subsequently, use them to calculate the key rate. 
The detailed derivation of this section is shown in Appendix \ref{Decoy State Analysis -- Yield Bounds} and \ref{Decoy State Analysis -- Error Yield Bounds}.

\subsection{KEY RATE}

The key rate for Passive MDI-QKD is given by  \cite{Wang2023a,Lo2012,Xu2013}
\begin{widetext}
\begin{equation}
	\begin{aligned}
		R& = P^A_Z P^B_Z
		\left[ \left\langle  P_1^A \right\rangle_{S_z}\left\langle  P_1^B \right\rangle_{S_z} Y_{11}^{Z, Lower, mixed} (1-h_2(e^{X, Upper, mixed}_{11}))
		- f \left\langle  Q^{AB}_Z \right\rangle_{S_z}  h_2(
		\left\langle Q^{AB}_Z E^{AB}_Z\right\rangle_{S_z}/
		\left\langle  Q^{AB}_Z \right\rangle_{S_z}) \right],
	\end{aligned}
\end{equation}
\end{widetext}

where $P^{A/B}_Z$ represents the probability for Alice/Bob to choose the key-generation basis, Z-basis. 
$\left\langle  P_1^{A/B} \right\rangle_{S_z}$ denotes the average probability for Alice/Bob to send a single-photon state in the key generation region.
$Y_{11}^{Z, Lower, mixed}$ is the lower bound of single-photon states in X basis, found from linear programming, and $e^{X, Upper, mixed}_{11}$ is the upper bound of error rate in X basis. 
$Q^{AB}_Z$ and $E^{AB}_Z $ are Z-basis Gain and QBER.

\begin{figure}[h]
	\centering
	\includegraphics[height=7cm]{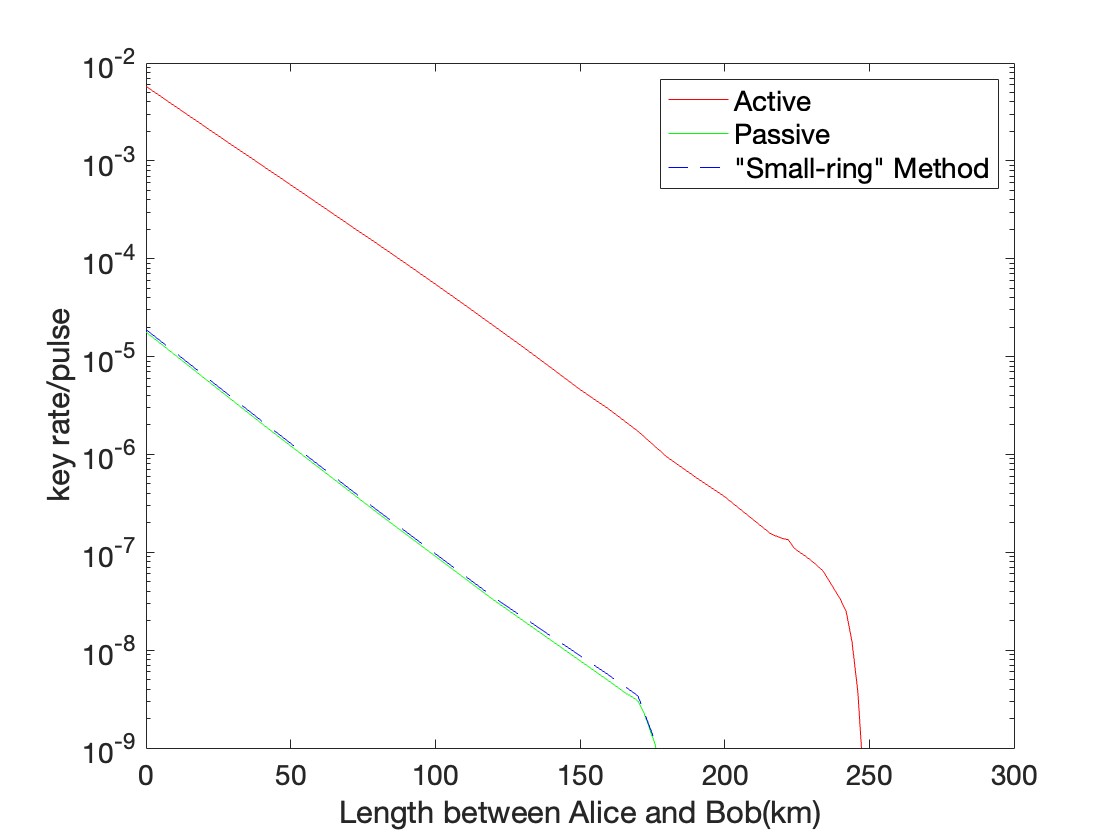}
	\caption{
		Key rate of passive MDI-QKD (green) and active MDI-QKD (red), along with the key rate of the newly proposed method (dashed blue, see Discussion Section).
		There is a misalignment of $1\%$ (Alice and Bob $0.5\%$ each) on the X-Z plane, and 3 decoy settings are used in both passive and active schemes.
		In the passive cases, $\Delta_Z$ is optimized within the range $[0.001, 0.05]$ and $t_3 $ is optimized within the range $[0.001, 0.99]$.
		The other parameters used were fixed, $\Delta_{XY}= 0.005, \Delta_{\phi}= 0.005, t_{1} = 0.005/u_{max},t_{2} = 0.05/u_{max}$.
		In the active case, the 3 intensities $[
		\mu, \nu, \omega]$ were used, all three intensities were optimized within the range $[0, 1.0]$.
		In both cases, the channel loss coefficient used is $0.2 \text{dB km}^{-1}$ and the detectors' dark count used is $10^{-6}$.
		The passive MDI-QKD's key rate is reasonable.
		It is about two to three orders of magnitudes lower than the active one, however, possesses the advantage of the removal of all source modulator side channels.
		In the "small-ring" method key rate calculation, the three decoy state intensities used are the same as the passive one.
		We see an improvement of about $6\% $.
	}
	\label{result}
\end{figure}

\section{SIMULATED RESULTS}

In this section, we show the simulated results of the key rate for asymptotic passive MDI-QKD. 
In Figure \ref{result}, we plot the key rate against communication distance for both an active and a passive MDI-QKD system. 
Both systems used 3 decoy settings.
We see that passive sources lower the key rate by two to three orders of magnitude than active MDI-QKD. 
This difference is mainly due to the double-sifting nature of passive MDI-QKD, where both Alice and Bob discarded many signals while doing post-selection. 
Passive MDI-QKD could reach a maximum communication distance of about 110km.
However, sacrificing some key rate performance, passive MDI-QKD allows the removal of side channels from both source modulators and detectors.

\section{DISCUSSIONS}

\begin{figure*}[t]
	\centering
	\includegraphics[height=8cm]{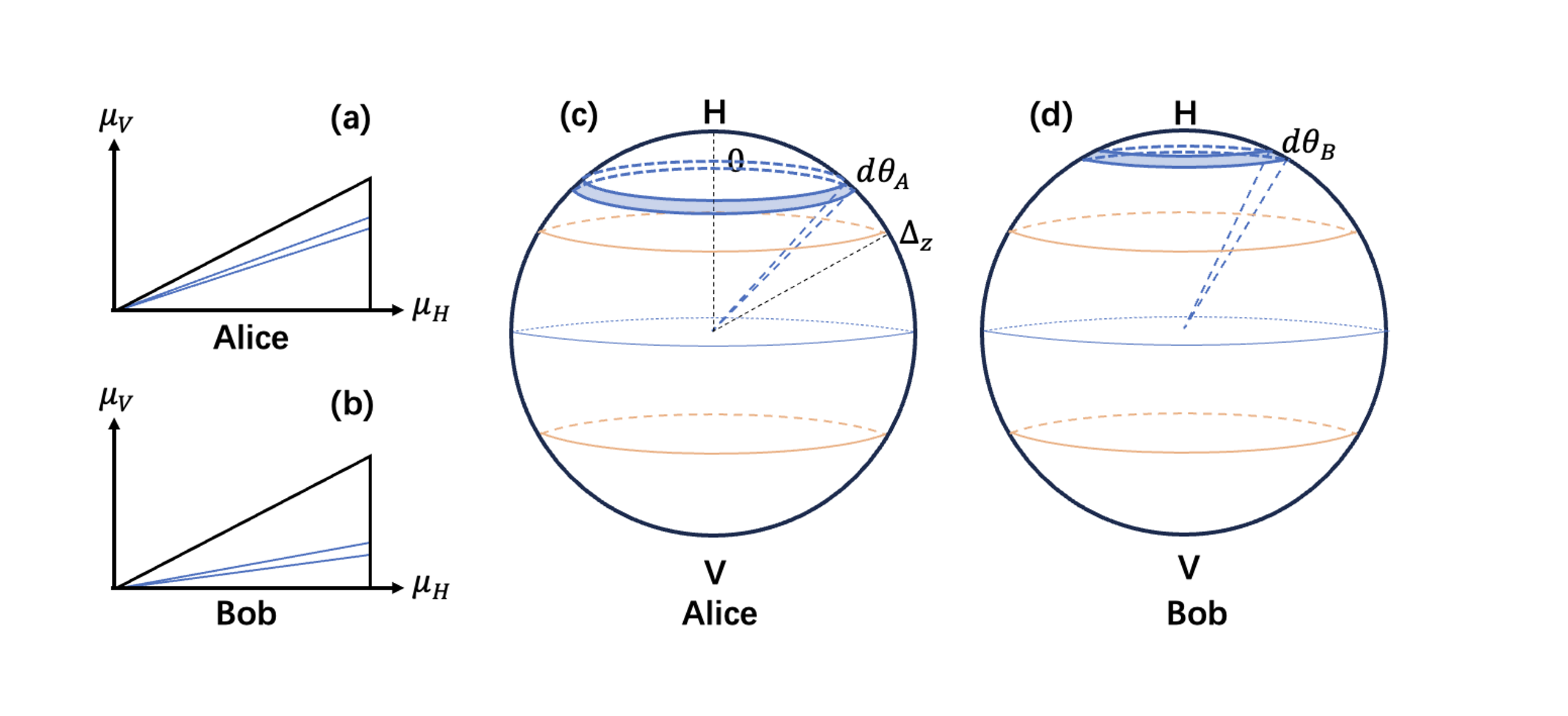}
	\caption{An additional method that could improve the key rate. 
		The key generation regions, $H$ and $V$, are subdivided into smaller rings (shaded regions between dark blue lines), and the key rate of each 'ring' is calculated and integrated together, see Equation \ref{improve}. 
		(a) and (b) shows the ring's corresponding regions in $\mu_H-\mu_V$ space for Alice and Bob, respectively.
		(c) and (d) shows the rings on the Bloch spheres for Alice and Bob, respectively.
		The rings are labeled by the polar angle, $\theta$, and have a thickness of $d \theta$. On Alice's Bloch sphere, the integration limits of Equation \ref{improve}, $0$, and $\Delta_Z$, are labeled. 
		Bob's integration limits are the same. 
	}
	\label{onion}
\end{figure*}

In this paper, we proposed a QKD scheme that implements the recently developed fully passive sources on MDI-QKD -- a passive MDI-QKD, that compromises some performance with the removal of all source modulator side channels. 
Recently, a fully passive TF-QKD \cite{Wang2023}was also proposed, which also could eliminate source modulator side channels.
However, our protocol is easier to implement because it does not require a global phase reference. 
The key rate of passive MDI-QKD is two to three order-of-magnitude lower than active MDI-QKD (see Figure \ref{result}), mainly due to the double-sifting nature of the passive sources. 
Further efforts could be made to tackle the sifting problem, which could improve the key rate performance significantly. 
Meanwhile, an experimental realization of the proposed scheme is also planned in our subsequent research.

We propose one more method that could improve the key rate.
In this method, we divide the key generation regions ($H$ and $V$ regions) into small rings, which can be in principle infinitesimally small, and one can calculate the key rate of each ring.
In other words, each ring is represented by a polar angle $\theta_{HV}$, hence the key rate of each ring is $R(\theta_{HV})$. 
One then integrates the key rates

\begin{equation}
	\begin{aligned}
		R_{improved} =  &\int_{0}^{\Delta_Z}\int_{0}^{\Delta_Z} R(\theta_{HVA},\theta_{HVB})
		\\&
		\times
		p(\theta_{HVA})p(\theta_{HVB}) d\theta_{HVA}d\theta_{HVB},
		\label{improve}
	\end{aligned}
\end{equation}
where the limits of the integration are the key generation region and $p(\theta_{HV})$ is the marginal of the intensity probability distribution (Equation \ref{keydistribution}) in terms of the polar angle.
Intuitively, one might obtain an improved key rate because we make use of more information.
Some 'rings' of too large of a $\theta_{HV}$ will have a too-big QBER, thus having a zero key rate. 
Using this method, those ‘zero key rate regions’ would be excluded from the overall key rate calculation.
However, while calculating the key rate of one bulk region, those regions would increase QBER while making no contribution to the key rate.
Specifically, the privacy amplification term remains the same because the lower/upper bounds of single-photon yield/error yield(from decoy state analysis) remains the same,
so summing the privacy amplification term comes down to just summing the number of single photons  (which is multiplied to a  constant $Y_{11}^{Z, Lower, mixed} [1-h_2(e^{X, Upper, mixed}_{11})]$), the total number of which is independent of the binning method.
However, less error correction is required for the small-ring method, because the binary entropy function is convex. Due to Jensen’s inequality \cite{jensen_2018_2371297}, the average of the entropy functions of respective errors in each ring is always less than the entropy function of the average error,$\left\langle h_2( E) \right\rangle < h_2(\left\langle E \right\rangle )$, thus making the small ring method advantageous.

The key rate using this new method is shown in Figure \ref{result} (blue dashed line). 
Comparing it to the passive MDI-QKD key rate, we observe an improvement of about $6\%$.
This method is  applicable to all potocols that use the fully passive sources, like fully passive BB84 \cite{Wang2023a}, not just to MDI-QKD.

\begin{acknowledgments}
We thank H.F Chau, R Wang, C Hu and X Lin for their helpful discussions. 
This project is financially supported by the University of Hong Kong start-up grant and NSERC. H.K. Lo also acknowledges support from MITACS and Innovative Solutions Canada. W.W. acknowledges support from the Hong Kong RGC General Research Fund (GRF) and the University of Hong Kong Seed Fund for Basic Research for New Staff.
\end{acknowledgments}

~

\noindent
\textit{Note added: During the preparation of this work, it came to our knowledge that another work on fully passive MDI-QKD is being prepared \cite{Wang2023b}, which is independently completed from our work. Our works are simultaneously posted on the preprint server \cite{Wang2023b, Li2023} .}

\appendix

\section{MDI-QKD \cite{Lo2012}}
\label{MDI-QKD}
In 2012, a novel QKD scheme, termed as MDI-QKD, was introduced to remove all detector side channels \cite{Lo2012}. 

Alice and Bob individually and randomly prepare one of the BB84 states using WCP sources, transferring them to an untrusted third party Charlie. 
Intensity modulators are applied to perform decoy state analysis \cite{Lo2005,Hwang2003,Wang2005}.
Charlie carries out a Bell state measurement (BSM), resulting in one of the Bell states as the outcome. 
Specifically, Charlie could arrange the measurement setup as depicted in Figure \ref{setup} (Charlie's side), where the two signals received from Alice and Bob could be interfered with at a 50:50 beam splitter (BS), and at each end of the BS, a polarization beam splitter (PBS) could project the signal into either a horizontal or vertical polarization state. 
At the end of each path, there is a single-photon detector (SPD).
Charlie publicly broadcasts the successful detection results. 
If precisely two detectors' clicks are observed, it's counted as one successful BSM, yielding two different types of Bell states. 
If detectors 1H and 2V or 1V and 2H click, it signifies a projection on the singlet Bell state $\left| \psi^- \right\rangle$. 
If detectors 1H and 1V, or 2H and 2V click, it indicates a projection on the triplet Bell state, $\left| \psi^+ \right\rangle$ \cite{Lo2012}.

When Alice and Bob receive Charlie's measurement results, they keep only those light signals that correspond to the successful measurements. 
Then they post-select those signals where Alice and Bob both use the same basis \cite{Lo2012}. 
The details of the implementation of MDI-QKD can be found in   \cite{Lo2012}.

Two bases are utilized separately: the rectilinear basis is employed for key generation, and the diagonal basis is used for testing. 
Consequently, the gain and QBER for the two bases should be calculated independently. The key rate is given by \cite{Lo2012}
\begin{equation} 
	R = Q^{11}_{rect} (1 - H(e^{11}_{diag})) - Q_{rect}f(E_{rect}) H(E_{rect})
\end{equation}
where $H(x) = -x log(x) - (1-x)log(1-x)$ is the Shannon entropy function.

\section{Decoy State Analysis -- Linear Programming Construction \cite{Wang2023a}}
\label{Decoy State Analysis -- Linear Programming Construction}

In passive MDI-QKD, the Gain/ErrorGain relations 
\begin{equation}
	\begin{aligned}
		\langle Q\rangle_{S_i^A S_j^B}
		&=\sum_{n,m=0}^{\infty}
		\left\langle P_n^A P_m^B 
		Y_{nm}\right\rangle_{S_i^A S_j^B}
		\\
		\langle Q E\rangle_{S_i^A S_j^B}&
		=\sum_{n,m=0}^{\infty}\left\langle P_n^A P_m^B  e_{nm} Y_{nm}\right\rangle_{S_i^A S_j^B}
	\end{aligned}
\end{equation}
where $S_i^A$ and $S_j^B$ are post-selection regions chosen by Alice and Bob, $P_n^{A/B}$ is Possion distribution, and are functions of $(\mu_H^{A/B}, \mu_V^{A/B})$, and  $Y_{nm}$ or $e_{nm} Y_{nm}$ are functions of 
$(\mu_H^{A}, \mu_V^{A}, \phi_{HV}^{A}, \mu_H^{B}, \mu_V^{B}, \phi_{HV}^{B})$, observables of both Alice and Bob. 
We can rewrite the coordinate into polar coordinates, $(\mu^A_H,\mu^A_H)$ into $(r^A, \theta^A)$ and similarly for Bob. 
We denote $\phi_{HV}^A$ into $\phi^A$. 

We now can write explicitly
\begin{equation}
	\begin{aligned}
		&\left\langle P_n^A P_m^B Y_{nm}\right\rangle_{S_i^A S_j^B}
		=\frac{1}{P_{S_i^A S_j^B}^{\mu^A, \mu^B, \phi^A, \phi^B}} 
		\\
		&\times
		\iiint \iiint_{S_i^A S_j^B} 
		p^A_\mu\left(r^A, \theta^A\right)
		p^B_\mu\left(r^B, \theta^B\right)
		p^A_\phi\left(\phi^A\right)
		p^B_\phi\left(\phi^B\right)    \\
		&\times
		P_n^A(r^A, \theta^A)P_m^B(r^B, \theta^B)
		\\
		&\times
		Y_{nm}(\theta^A, \phi^A, \theta^B, \phi^B) r^A r^B dr^A dr^B d\theta^A d\theta^B d\phi^A d\phi^B
	\end{aligned}
\end{equation}

\begin{equation}
	\begin{aligned}
		&\left\langle P_n^A P_m^B  e_{nm} Y_{nm}\right\rangle_{S_i^A S_j^B}
		=\frac{1}{P_{S_i^A S_j^B}^{\mu^A, \mu^B, \phi^A, \phi^B}} 
		\\
		&\times
		\iiint \iiint_{S_i^A S_j^B} 
		p^A_\mu\left(r^A, \theta^A\right)
		p^B_\mu\left(r^B, \theta^B\right)
		p^A_\phi\left(\phi^A\right)
		p^B_\phi\left(\phi^B\right)    \\
		&\times
		P_n^A(r^A, \theta^A)P_m^B(r^B, \theta^B)
		\\
		&\times
		e_{nm}Y_{nm}(\theta^A, \phi^A, \theta^B, \phi^B) r^A r^B dr^A dr^B d\theta^A d\theta^B d\phi^A d\phi^B
	\end{aligned}
\end{equation}

Note that here, $Y_{nm}$ depends on $\theta$ and $\phi$, the polarization. 
Since $Y_{nm}$ and $e_{nm}Y_{nm}$ are in similar form, from here, we will just focus on the calculation of $Y_{nm}$.
The other one follows the same procedure.
The following procedure is largely inspired by \cite{Wang2023a}.

One could take out the integral of $\phi^A$ and $\phi^B$ part,
\begin{equation}
	\begin{aligned}
		&\left\langle P_n^A P_m^B Y_{nm}\right\rangle_{S_i^A S_j^B}
		 =\frac{1}{P_{S_i^A S_j^B}} 
		\iint \iint_{S_i^A S_j^B} 
		p^A_\mu\left(r^A, \theta^A\right)
				\\
		&\times
		p^B_\mu\left(r^B, \theta^B\right)    
		\\
		&\times
		P_n^A(r^A, \theta^A)P_m^B(r^B, \theta^B)
		\\
		&\times
		Y_{nm}(\theta^A, \theta^B) r^A r^B dr^A dr^B d\theta^A d\theta^B 
	\end{aligned}
\end{equation}
here, we partially integrate $Y_{nm}$ over $\phi^A$, $\phi^B$
\begin{equation}
	\begin{aligned}
	Y_{nm}(\theta^A, \theta^B) = &
	\frac{1}{P_{S_i^A S_j^B}^{\phi^A, \phi^B}}
\iint_{\phi^A, \phi^B} 
p^A_\phi\left(\phi^A\right)
p^B_\phi\left(\phi^B\right)
\\& \times
Y_{nm}(\theta^A, \phi^A, \theta^B, \phi^B)
d\phi^A d\phi^B
	\end{aligned}
\end{equation}

From the above, we can rearrange the integrals by collecting terms related to $r^{A/B}$.
From above, 
\begin{equation}
	\begin{aligned}
	&	\left\langle P_n^A P_m^B Y_{nm}\right\rangle_{S_i^A S_j^B}
		 =\frac{1}{P_{S_i^A S_j^B}} 
		\iint \iint_{S_i^A S_j^B} 
		p^A_\mu\left(r^A, \theta^A\right)
				\\
		&\times
		p^B_\mu\left(r^B, \theta^B\right)    \\
		&\times
		P_n^A(r^A, \theta^A)
		P_m^B(r^B, \theta^B)    \\
		&\times
		Y_{nm}(\theta^A, \theta^B) r^A r^B dr^A dr^B d\theta^A d\theta^B 
		\\
		&=
		\iint_{\theta^{A/B}} 
		\left( \frac{
			\int_{r^A(\theta^A)}
			p^A_\mu\left(r^A, \theta^A\right)
			P_n^A(r^A, \theta^A)
			r^A dr^A }{P_{S_i^A}}
		\right)    \\
		&\times
		\left(\frac{
			\int_{r^B(\theta^B)}
			p^B_\mu\left(r^B, \theta^B\right)
			P_m^B(r^B, \theta^B)
			r^B dr^B }   {P_{S_i^A}}
		\right)    \\
		&\times
		Y_{nm}(\theta^A, \theta^B)  d\theta^A d\theta^B 
		\\
		& = 
		\left\langle P_n^A \right\rangle_{S_i^A}
		\left\langle P_n^B \right\rangle_{S_j^B}
		\iint_{\theta^{A/B}} 
		\frac{p_{\theta^A, n, S_i^A}(\theta^A)}
		{\left\langle P_n^A \right\rangle_{S_i^A}}
		\frac{p_{\theta^B, m, S_j^B}(\theta^A)}
		{\left\langle P_n^B \right\rangle_{S_j^B}}    \\
		&\times
		Y_{nm}(\theta^A, \theta^B)  d\theta^A d\theta^B 
		\\
		& = 
		\left\langle P_n^A \right\rangle_{S_i^A}
		\left\langle P_n^B \right\rangle_{S_j^B}
		\times
		Y^{mixed}_{nm, S_i S_j}
	\end{aligned}
\end{equation}
where 
\begin{equation}
	P_{S_i^A} = \iint_{S_i^A}     p^A_\mu\left(r^A, \theta^A\right)
	r^A  dr^A  d\theta^A 
\end{equation}
and 
\begin{equation}
	P_{S_i^A} P_{S_J^B} =  P_{S_i^A S_j^B}
\end{equation}
and we have replaced 

\begin{equation}
	p_{\theta^A, n, S_i^A}(\theta^A)
	=
	\frac{
		\int_{r^A(\theta^A)}
		p^A_\mu\left(r^A, \theta^A\right)
		P_n^A(r^A, \theta^A)
		r^A dr^A }{P_{S_i^A}}
\end{equation}
and finally we replaced \cite{Wang2023a}

\begin{equation}
\begin{aligned}
	Y^{mixed}_{nm, S_i S_j}
=&
\iint_{\theta^{A/B}} 
\frac{p_{\theta^A, n, S_i^A}(\theta^A)}
{\left\langle P_n^A \right\rangle_{S_i^A}}
\frac{p_{\theta^B, m, S_j^B}(\theta^A)}
{\left\langle P_n^B \right\rangle_{S_j^B}}
\\&\times
Y_{nm}(\theta^A, \theta^B)  d\theta^A d\theta^B 
\end{aligned}
\end{equation}

Up to this point, we have effectively decoupled the functions. 

\begin{equation}
	\begin{aligned}
		\left\langle P_n^A P_m^B Y_{nm}\right\rangle_{S_i^A S_j^B}
		& =
		\left\langle P_n^A \right\rangle_{S_i^A}
		\left\langle P_n^B \right\rangle_{S_j^B}
		\times
		Y^{mixed}_{nm, S_i S_j}.
	\end{aligned}
\end{equation}
However, linear programming is not yet applicable because $Y^{mixed}_{nm,S_i S_j}$ depends on the decoy regions $S_{i/j}$, therefore, $Y^{mixed}_{nm,S_i S_j}$ is not constant, so one could not bound them using linear programming. 

To solve the problem, 
we cleverly construct a decoy setting so that the yield $Y^{mixed}_{nm}(\theta)$ is independent of the decoy settings, namely, it is consistent across all decoy settings so that a normal linear program can be used \cite{Wang2023a}. 
The key is to let Alice and Bob perform an additional post-selection to 'shape' the probability distribution into the desired form \cite{Wang2023a}. 

More Concretely, for both Alice and Bob, they perform an additional post-selection to make the overall probability distribution to $p^{\prime}_{\mu} = C e^{r ( sin \theta + cos\theta) }$ \cite{Wang2023a}.
In this way, the exponential part in the Poisson term can be canceled out. 
Finally we will have \cite{Wang2023a}
\begin{widetext}
\begin{equation}
	Y^{mixed}_{nm} = 
	\frac{
		\iint_{\theta^{A/B}} 
		(sin \theta^A + cos \theta^A )^n (sin \theta^B + cos \theta^B )^m  Y_{nm}(\theta^A, \theta^B) d\theta^A d\theta^B 
	}{
		\iint_{\theta^{A/B}} 
		(sin \theta^A + cos \theta^A )^n (sin \theta^B + cos \theta^B )^m   d\theta^A d\theta^B 
	}
\end{equation}
\end{widetext}

Meanwhile, they also need to constrain the decoy regions to sector-shaped regions, see Figure \ref{post}. 

Now, the linear program can be written as 
\begin{equation}
	\langle Q\rangle_{S_i^A S_j^B}
	= 
	\left\langle P_n^A \right\rangle_{S_i^A}
	\left\langle P_n^B \right\rangle_{S_j^B}
	\times
	Y^{mixed}_{nm}
\end{equation}
and a similar form for the error yield.
\begin{equation}
	\langle QE\rangle_{S_i^A S_j^B}
	= 
	\left\langle P_n^A \right\rangle_{S_i^A}
	\left\langle P_n^B \right\rangle_{S_j^B}
	\times
	e_{nm}Y^{mixed}_{nm}
\end{equation}

\section{Decoy State Analysis -- Yield Bounds \cite{Wang2023a}}
\label{Decoy State Analysis -- Yield Bounds}

We argue that the lower bounds of the 'mixed' single-photon yield are actually that of the 'perfectly encoding' yield, mathematically, 
\begin{equation}
	Y^{mixed, Lower}_{11} \le Y^{mixed}_{11} = Y_{11}^{perfect}
	.\end{equation}

In passive MDI-QKD, we need both Alice and Bob to get an H state or a V state, label the coordinates with $\theta_{A/B}$ and $\phi_{A/B}$, respectively \cite{Wang2023a}. 
We should have the mixed state 
\begin{equation}
	\begin{aligned}
	\rho^{mixed} = &
\left| H^{\prime} H^{\prime} \right\rangle \left\langle H^{\prime} H^{\prime} \right| +
\left| H^{\prime}V^{\prime} \right\rangle \left\langle H^{\prime}V^{\prime} \right|
\\ &+
\left| V^{\prime} H^{\prime} \right\rangle \left\langle V^{\prime} H^{\prime} \right| +
\left| V^{\prime}V^{\prime} \right\rangle \left\langle V^{\prime}V^{\prime} \right|
	\end{aligned}
\end{equation}

We simplify that $\phi_A = 0 $ and $\phi_B= \phi$, the global phase does not really matter, and we set the global phase to Alice's phase so that $\phi_A = 0$.

Start from the first term, and the others just follow similar construction,
\begin{equation}
	\begin{aligned}
		\left| H^{\prime} H^{\prime} \right\rangle & = \left| H^{\prime}\right\rangle_A
		\otimes
		\left| H^{\prime}  \right\rangle_B
		\\
		&=  \left(
		cos(\frac{\theta_A}{2})\left| H \right\rangle_A 
		+ 
		sin(\frac{\theta_A}{2})\left| V \right\rangle_A\right)
		\\ &
		\otimes  \left(
		cos(\frac{\theta_B}{2})\left| H \right\rangle_B + 
		e^{i \phi} sin(\frac{\theta_B}{2})\left| V \right\rangle_B\right)
	\end{aligned}
\end{equation}
Therefore, the density matrix 
\begin{widetext}
\begin{equation}
	\begin{aligned}
		\left| H^{\prime} H^{\prime} \right\rangle \left\langle H^{\prime} H^{\prime} \right| 
		&=
		\begin{pmatrix}
			cos(\frac{\theta_A}{2})cos(\frac{\theta_B}{2}) & e^{-i \phi} cos(\frac{\theta_A}{2})sin(\frac{\theta_B}{2}) & sin(\frac{\theta_A}{2})cos(\frac{\theta_B}{2}) & e^{-i \phi} sin(\frac{\theta_A}{2})sin(\frac{\theta_B}{2})\\
		\end{pmatrix}
		\\
		&
		\otimes
		\begin{pmatrix}
			cos(\frac{\theta_A}{2})cos(\frac{\theta_B}{2}) \\ e^{i \phi} cos(\frac{\theta_A}{2})sin(\frac{\theta_B}{2}) \\ sin(\frac{\theta_A}{2})cos(\frac{\theta_B}{2}) \\ e^{i \phi} sin(\frac{\theta_A}{2})sin(\frac{\theta_B}{2})\\
		\end{pmatrix}
	\end{aligned}
\end{equation}
\end{widetext}

Similarly, we can construct another basis
\begin{equation}
	\begin{aligned}
		\left| H^{\prime} V^{\prime} \right\rangle \left\langle H^{\prime} V^{\prime} \right| 
	\end{aligned}
\end{equation}

\begin{equation}
	\begin{aligned}
		\left| V^{\prime} H^{\prime} \right\rangle \left\langle V^{\prime} H^{\prime} \right| 
	\end{aligned}
\end{equation}

\begin{equation}
	\begin{aligned}
		\left| V^{\prime} V^{\prime} \right\rangle \left\langle V^{\prime} V^{\prime} \right| 
	\end{aligned}
\end{equation}
adding them all together, it is not hard to spot that the diagonal terms are all $1$, and non-diagonal terms are all $0$.
Hence, the whole matrix is a 4-dimensional identity matrix. 

\begin{equation}
	\begin{aligned}
		 \rho^{\prime} & = 
		\left| H^{\prime} H^{\prime} \right\rangle \left\langle H^{\prime} H^{\prime} \right| +
		\left| H^{\prime}V^{\prime} \right\rangle \left\langle H^{\prime}V^{\prime} \right| 
		\\ & +
		\left| V^{\prime} H^{\prime} \right\rangle \left\langle V^{\prime} H^{\prime} \right| +
		\left| V^{\prime}V^{\prime} \right\rangle \left\langle V^{\prime}V^{\prime} \right|
		\\
		& = \begin{pmatrix}
			1 &  &  &  \\
			& 1 &  &  \\
			&  & 1 &  \\
			&  &  & 1 \\
		\end{pmatrix}= I
	\end{aligned}
\end{equation}
Note that we have used the trigonometry identity $sin^2(x)+cos^2(x) = 1$.
Up to now, we have shown that the 'mixed' state is actually a fully-mixed state, $I$, so that it is equivalent to the 'perfect encoding state', which is just 
\begin{equation}
	\left| H H \right\rangle \left\langle H H \right| +
	\left| HV \right\rangle \left\langle HV \right|+
	\left| V H \right\rangle \left\langle V H \right| +
	\left| VV \right\rangle \left\langle VV \right| = I
\end{equation}
So, the 'mixed' state and the 'perfect' state, $\rho^{perfect}$, should share a lower bound. 
Hence, we could use the lower bound of $ Y^{mixed}_{nm}$, denote it $ Y^{mixed, Lower}_{nm}$, as the lower bound of the 'perfectly encoded' state. 
\begin{equation}
	Y^{mixed, Lower}_{11} \le Y^{mixed}_{11} = Y_{11}^{perfect}
\end{equation}

\section{Decoy State Analysis -- Error Yield Bounds \cite{Wang2023a}}
\label{Decoy State Analysis -- Error Yield Bounds}

We only consider the $HH$ state here(both Alice and Bob send an $H$ state) \cite{Wang2023a}.
We would like to compute
\begin{equation}
	\begin{aligned}
	\rho^{HH} & = 
\left| H_1 H_1^{\perp}  \right\rangle \left\langle H_1 H_1^{\perp}  \right| +
\left| H_1 H_2^{\perp}  \right\rangle \left\langle H_1 H_2^{\perp}   \right| 
\\ & +
\left| H_2 H_1^{\perp}  \right\rangle \left\langle H_2 H_1^{\perp}  \right| +
\left| H_2 H_2^{\perp} \right\rangle \left\langle H_2 H_2^{\perp}  \right|
	\end{aligned}
\end{equation}
where $ \left| H_1 H_1^{\perp}  \right\rangle$ is $ \left| H_1  \right\rangle \otimes  \left| H_1^{\perp}  \right \rangle$, and we define $ \left| H_1  \right\rangle$ is just a 'polarized' $H$ state, which has the coordinate $(\theta_A , \phi_A)$ on the Bloch sphere. 
The state $ \left| H_1^{\perp}\right\rangle$ is the 'pair state' with coordinate $(\theta_A , \phi_A+\pi)$. 
Similarly, states $ \left| H_2  \right\rangle$ and $ \left| H_2^{\perp}\right\rangle$ are Bob's analog states, with coordinates $(\theta_B, \phi_B)$ and $(\theta_B, \phi_B+\pi)$. respectively \cite{Wang2023a}. 
We again offset the phases so that Alice has no phase, and Bob simply has a $\phi$ phase. 

Therefore, we can write down each of the four states in terms of 
\begin{equation}
	\left| H_1  \right\rangle =  
	cos(\frac{\theta_A}{2})\left| H \right\rangle_A 
	+ 
	sin(\frac{\theta_A}{2})\left| V \right\rangle_A
\end{equation}
The remaining 3 states can be written down in a similar form. 

Hence, we could compute the terms, 
$\left| H_1 H_1^{\perp}  \right\rangle \left\langle H_1 H_1^{\perp}  \right| $,$
\left| H_1 H_2^{\perp}  \right\rangle \left\langle H_1 H_2^{\perp}   \right|$, $
\left| H_2 H_1^{\perp}  \right\rangle \left\langle H_2 H_1^{\perp}  \right|$,$
\left| H_2 H_2^{\perp} \right\rangle \left\langle H_2 H_2^{\perp}  \right| $ individually. 
They should each be a $4\times 4 $ matrix. 

Adding them up, we end up with a matrix
\begin{widetext}
\begin{equation}
	\rho = 
	\begin{pmatrix}
		cos^2(\frac{\theta_A}{2})cos^2(\frac{\theta_B}{2}) &  &  &  \\
		& cos^2(\frac{\theta_A}{2})sin^2(\frac{\theta_B}{2}) &  &  \\
		&  & sin^2(\frac{\theta_A}{2})cos^2(\frac{\theta_B}{2}) &  \\
		&  &  & sin^2(\frac{\theta_A}{2})sin^2(\frac{\theta_B}{2}) \\
	\end{pmatrix}
\end{equation}
\end{widetext}

We could rewrite the matrix into a combination of states:
\begin{equation}
	\begin{aligned}
		\rho  & = 
		(cos^2(\frac{\theta_A}{2})-sin^2(\frac{\theta_A}{2}))(cos^2(\frac{\theta_B}{2})sin^2(\frac{\theta_B}{2}))
		\begin{pmatrix}
			1 &  &  &  \\
			& 0 &  &  \\
			&  & 0 &  \\
			&  &  & 0 \\
		\end{pmatrix}
		\\
		& + 
		(cos^2(\frac{\theta_A}{2})-sin^2(\frac{\theta_A}{2}))sin^2(\frac{\theta_B}{2})
		\begin{pmatrix}
			1 &  &  &  \\
			& 1 &  &  \\
			&  & 0 &  \\
			&  &  & 0 \\
		\end{pmatrix}
		\\
		&+
		(cos^2(\frac{\theta_B}{2})-sin^2(\frac{\theta_B}{2}))sin^2(\frac{\theta_A}{2})
		\begin{pmatrix}
			1 &  &  &  \\
			& 0 &  &  \\
			&  & 1 &  \\
			&  &  & 0 \\
		\end{pmatrix}
		\\
		& +
		sin^2(\frac{\theta_A}{2})sin^2(\frac{\theta_B}{2})
		\begin{pmatrix}
			1 &  &  &  \\
			& 1 &  &  \\
			&  & 1 &  \\
			&  &  & 1 \\
		\end{pmatrix},
	\end{aligned}
	\label{eq.big}
\end{equation}

\begin{equation}
	\begin{aligned}
		\rho  & = 
		(cos^2(\frac{\theta_A}{2})-sin^2(\frac{\theta_A}{2}))(cos^2(\frac{\theta_B}{2})sin^2(\frac{\theta_B}{2})) \\& \times
		( \left| HH \right\rangle \left\langle HH \right| )
		\\
		&+
		(cos^2(\frac{\theta_A}{2})-sin^2(\frac{\theta_A}{2}))sin^2(\frac{\theta_B}{2}) 
		( \left| H \right\rangle \left\langle H \right|)_A \otimes I_B )
		\\
		\\
		&+
		(cos^2(\frac{\theta_B}{2})-sin^2(\frac{\theta_B}{2}))sin^2(\frac{\theta_A}{2})
		(    I_A \otimes     \left| H \right\rangle \left\langle H \right|)_B)
		\\
		&+
		sin^2(\frac{\theta_A}{2})sin^2(\frac{\theta_B}{2})
		(I_A \otimes I_B).
	\end{aligned}
	\label{eq.big}
\end{equation}

We can notice that the last 3 terms all have a QBER of $50\%$, and we also notice that the first term is the 'perfect' encoding state, which must have a smaller QBER than the 'mixed' state \cite{Wang2023a}.

For H state alone(for both Alice and Bob), if we integrate over all possible pair of polarized state \cite{Wang2023a}, each has a density matrix like Equation \ref{eq.big}, the overall density matrix for the $HH$ scenario (when Alice has an H state and Bob has an H state) would be the sum of some sort of mixture(the latter 3 terms in  Equation \ref{eq.big}) and the perfectly encoding state(the first term in  Equation \ref{eq.big}). 
The mixture terms have a QBER of $50 \%$, 
Therefore, we could conclude that 
\begin{equation}
	e_{11}Y_{11}^{mixed} \ge e_{11}Y_{11}^{perfect},
\end{equation}
similarly, for $VV$ , $HV$ and $VH$ states. 
A similar argument could be proved for $++$, $--$, $+-$ and $-+$ state in $X$ basis.

Therefore, we could make the following conclusion:
the upper bound of the mixed state, $e_{nm}Y_{nm}^{mixed, Upper}$, which can be obtained from linear programming, is indeed the upper bound of the 'perfect' state, $eY^{perfect}$.

\begin{equation}
	e_{11}Y_{11}^{mixed, Upper} \ge e_{11}Y_{11}^{ mixed}  \ge   e_{11}Y_{11}^{perfect}.
\end{equation}

\section{Channel Model}
\label{Channel Model}

The signals immediately coming out from Alice's fully passive source are described by four DOFs, $ (\mu_A, \theta_{HVA}, \phi_{HVA}, \phi_{global, A})$, and a similar set for Bob. 
The signals can be slightly polarized after they travel through the channel. 
Writing the Bloch sphere state in terms of coordinates \cite{Wang2023a}

\begin{equation}
	\Vec{s}_A = (sin \theta_{HVA} cos \phi_{HVA}, sin \theta_{HVA} sin \phi_{HVA} ,  cos \theta_{HVA}).
\end{equation}

This polarization rotation is described by the Rodrigues' formula \cite{Friedberg2022}

\begin{equation}
	\Vec{s'}_A = cos \alpha \Vec{r} + sin \alpha (\Vec{s} \times \Vec{r})+(\Vec{s} \cdot \Vec{r}) (1-cos\alpha) \Vec{s},
\end{equation}

where $\alpha$ is the rotation angle, and $\Vec{r}$ is the rotation axis (unit length). 
We then could convert the coordinate back into Bloch sphere coordinate $\{\mu'_A,\theta_{HVA}', \phi_{HVA}', \phi_{GA}'  \}$, where the superscripts denote post-rotation. 
It is noteworthy that the global phase is also rotated, however, does not affect the physics -- it will eventually be integrated over.

The former two post-rotation parameters for Alice $(\mu_A,\theta_{HVA}'  )$ can be reconverted into H and V leg intensities, $(\mu_{HA}',\mu_{VA}')$. Considering the channel loss and detector efficiency, the intensity arriving at the detectors becomes $\mu_{HA}' \rightarrow \mu_{HA}' \eta = \mu_{HA}'\eta_L \eta_d = \mu_{HA}' 10^{-\alpha L/10} \eta_d$, where $\alpha$ is the loss coefficient, and typical is about $0.2\text{dB}/ \text{km}$, $\eta_d$ is the detector efficiency.
The other intensities follow the same format.

The post-rotation signals will travel through the channels and arrive at the 50:50 BS for interference.
The interference can be partitioned into H and V components, each adhering to the interference formula, 
\begin{equation}
	\begin{aligned}
		&\left|\sqrt{\mu_{1}} e^{i \phi_{1}}\right\rangle_{a}\left|\sqrt{\mu_{2}} e^{i \phi_{2}}\right\rangle_{b} \rightarrow 
		\\
		&\left|\sqrt{\mu_{1} / 2} e^{i \phi_{1}}+i \sqrt{\mu_{2} / 2} e^{i \phi_{2}}\right\rangle_{c}\left|i \sqrt{\mu_{1} / 2} e^{i \phi_{1}}+\sqrt{\mu_{2} / 2} e^{i \phi_{2}}\right\rangle_{d}
	\end{aligned}
\end{equation}

we can derive the output intensity for leg c as 

\begin{equation}
	\mu_1/2 + \mu_2/2 - \sqrt{\mu_1 \mu_2}\sin(\phi)
\end{equation}

where $\phi$ represents the phase difference between the two signals. Meanwhile, the output intensity for leg d is given by 

\begin{equation}
	\mu_1/2 + \mu_2/2 + \sqrt{\mu_1 \mu_2}\sin(\phi).
\end{equation}

The intensities applied here are $\mu_{HA}',\mu_{HB}'$ and $\mu_{VA}',\mu_{VB}'$ for the H and V interference, respectively. 
For the H leg, the phase difference used is $\phi_{HA}' - \phi_{HB}' =\phi_{R}'$. For the V leg, the phase difference is determined by $\phi_{VA}' - \phi_{VB}' = \phi_{R}' + \phi_{HVA}'- \phi_{HVB}'$. 
Using the formula, we can compute the intensities reaching detectors 3H and 4H from the H leg interference, as well as the 3V and 4V intensities from the V leg interference. 

Because the phase differences $\phi_{R}'$ and $\phi_{R}' + \phi_{HVA}'- \phi_{HVB}'$ are randomized, in order to calculate the average gain, one needs to integrate the phases from $0$ to $2 \pi$, this explains the phase integration in Equation \ref{eq.expectation}.
The phase integration should also be separate for H and V; however, the phase difference in V interference is also a function of $\phi_{R}'$. Therefore, to derive the average gain, it suffices to integrate $\phi_{R}'$ from $0$ to $2 \pi$.

\bibliography{Maintext}

\end{document}